%% file: main.tex
\documentclass[conference]{IEEEtran}

\usepackage{amsmath}
\usepackage{url}
\usepackage[pdftex]{graphicx}
\usepackage{balance} 
\usepackage{booktabs}
\usepackage{siunitx}
\usepackage{cleveref}
\usepackage[most]{tcolorbox}

\newtcbtheorem{Summary}{\bfseries Summary}{enhanced,drop shadow={black!50!white},
  coltitle=black,
  top=0.15in,
  attach boxed title to top left=
  {xshift=1.5em,yshift=-\tcboxedtitleheight/2},
  boxed title style={size=small,colback=white}
}{summary}

\newtcolorbox[auto counter]{summary}[1][]{title={\bfseries Hypothesis~#1},enhanced,drop shadow={black!50!white},
  coltitle=black,
  top=0.15in,
  enlarge top by=0.08in,
  attach boxed title to top left=
  {xshift=1.5em,yshift=-\tcboxedtitleheight/2},
  boxed title style={size=small,colback=white},}

\begin{document}

\title{How much does AI impact development speed?\\ An enterprise-based randomized controlled trial}

\author{\IEEEauthorblockN{Elise Paradis, Kate Grey, Quinn Madison, Daye Nam,\\ Andrew Macvean, Vahid Meimand, Nan Zhang, Ben Ferrari-Church, Satish Chandra}
\IEEEauthorblockA{Google\\\{eparadis, kategrey, qmadison, dayenam, amacvean, vahidzm, nanzh, ferrarichurch, chandrasatish\}@google.com}
}

\maketitle

\thispagestyle{plain}
\pagestyle{plain}

\begin{abstract}
How much does AI assistance impact developer productivity? To date, the software engineering literature has provided a range of answers, targeting a diversity of outcomes: from perceived productivity to speed on task and developer throughput. Our randomized controlled trial with 96 full-time Google software engineers contributes to this literature by sharing an estimate of the impact of three AI features on the time developers spent on a complex, enterprise-grade task. We found that AI significantly shortened the time developers spent on task. Our best estimate of the size of this effect, controlling for factors known to influence developer time on task, stands at about 21\%, although our confidence interval is large. We also found an interesting effect whereby developers who spend more hours on code-related activities per day were faster with AI. Product and future research considerations are discussed. In particular, we invite further research that explores the impact of AI at the ecosystem level and across multiple suites of AI-enhanced tools, since we cannot assume that the effect size obtained in our lab study will necessarily apply more broadly, or that the effect of AI found using internal Google tooling in the summer of 2024 will translate across tools and over time.
\end{abstract}

\IEEEpeerreviewmaketitle

\input{paper_sections/01_introduction}
\input{paper_sections/02_related_work}

\input{paper_sections/03_dev_at_google}
\input{paper_sections/04_research_approach}
\input{paper_sections/05_limitations}
\input{paper_sections/06_results}
\input{paper_sections/07_discussion_implications}
\input{paper_sections/08_conclusion}
\input{paper_sections/09_acknowledgements}

\newpage
\balance

\bibliographystyle{IEEEtran}
\bibliography{references}

\end{document}

%% file: paper_sections/01_introduction.tex
\section{Introduction}
Seven years after the rise of LLM architecture~\cite{vaswani2017} and two years after the start of the ``chatbot revolution''~\cite{cnn-2023-chatRevolution}, significant investments have been made to AI-enhanced products, including in the software developer space. Since the release of GitHub Copilot~\cite{copilot}, numerous developer tools offering code editing and generation support have been built for the general developer community~\cite{alphacode, codewhisperer, tabnine, cursor}.  Researchers and educators have also developed prototype tools to assist novice programmers and students.  Furthermore, substantial effort has been invested in building tools for internal use, such as at Meta~\cite{bader2021} and Google~\cite{garg2022, froemmgen2024, AIsoftEng}.

However, there is still much to investigate to answer how useful these tools are in helping developers, specifically in improving their productivity.
Truly understanding the productivity benefits of AI enhanced coding tools remains a nascent field. While some research has shown improvements in coding speed~\cite{peng2023}, developer throughput~\cite{cui2024}, and perceived productivity~\cite{ziegler2022}, more work must be done to validate these assertions across, for example, tasks, developer contexts, user groups, and more.

To date, very few estimates of the impact of AI-enhanced developer tools on time spent on task in an enterprise context have been published. One much-discussed study was a randomized controlled trial by Peng et al.~\cite{peng2023} ($n = 95$), which found a 56\% speed increase for developers using GitHub Copilot---an AI code assistant---, compared to those not using it. Another enterprise-specific estimate comes from a pooled analysis of three field experiments ($n = 4,867$) conducted by Cui et al.~\cite{cui2024}, where developers either had access to Copilot or did not have access to it in their daily activities. The authors found a 26\% increase in throughput (measured as an increase in the number of pull requests) for developers using Copilot. 

Although these studies provide valuable insights and help quantify the speed improvements offered by one AI-enhanced developer tool, gaps remain in estimating the overall impact of different coding assistants and across industry settings. Other shortcomings include missing nuance or understanding of the impact of AI tools by developer- or task-level contexts (e.g., \cite{cui2024}), limited tool and task complexity in experimental settings (e.g., using GitHub Classroom rather than a more naturalistic developer environment~\cite{peng2023}), and small sample sizes (e.g., only 32 participants in~\cite{nam2024}, 24 participants in~\cite{vaithilingam2022}, or 21 participants in~\cite{imai2022}). Altogether, the rapidly-changing status of AI-enhanced developer tools, combined with the partial portraits provided by current studies at this very early stage in the empirical study of AI tools in production, requires continued inquiry.

In this paper, we complement this recent literature by providing an estimate for the impact of AI in an \textit{enterprise} context (as per~\cite{cui2024}) and applying it to \textit{speed on task} (as per~\cite{peng2023}). To simulate the enterprise context in the study, we designed a task covering multiple aspects of software development, from writing and editing code, to updating build files and to testing, within our proprietary internal infrastructure at Google. We aimed to answer the following questions specifically for our internal developer tools at Google:

\begin{itemize}
    \item RQ1: What impact does AI have on time spent completing an enterprise-grade development task?
    \item RQ2: How do developer and task characteristics influence our estimates of the impact of AI assistance on time spent on task?
    \item RQ3: How do developer and task characteristics interact with the use of AI to accelerate or slow down certain developers and not others?
\end{itemize}

Providing a robust estimate for the impact of AI-enhanced tools on development speed is critical to the long term adoption and success of these tools across the industry. Continued investment in, and adoption of, these tools is dependent not only on how developers feel about them; it is critical to be able to evaluate their business impact in terms of greater output or time gains for the organization. The second and third research questions unlock important new understandings around how to design and develop AI-enhanced developer tools in a product and user-centric manner. When we understand our users, we can better cater to their needs.

In this study, we ran a controlled trial with 96 Google software engineers who were randomly assigned either to use (experimental condition) or not use (control group) three AI-enhanced features for code (AI Code Completion, Smart Paste, and Natural Language to Code; see \Cref{sec:atGoogle} for more details) to complete an enterprise-grade task. We analyzed data using t-tests and linear regressions on time on task data to evaluate the impact of AI on speed on task. We ascertained the robustness of our estimate using multivariate regressions based on a theoretical framework~\cite{varpio2020}. 

Then, we answered our research questions by testing hypotheses we built based on the theoretical framework and the literature. For RQ1, we tested, ``H1: Participants randomly assigned to the AI condition will  spend  less  time  on  task'', and for RQ2, we tested ``H2:  Controlling  for  developer-level  and  task-level  factors,  participants  randomly  assigned  to  the  AI  condition will  still  spend  less  time  on  task.'' Finally, for RQ3, we tested hypotheses 3-5, ``There will be a negative and significant interaction effect  between  the  experimental  condition  and  average daily  hours  spent  coding (H3), seniority (H4), and the frequency with which developers use AI coding tools (H5)''.

Our study shows that developers who used AI were about 21\% faster than those who did not, controlling for other factors. We also found that more senior developers and developers who code more hours per day were significantly faster on the task than those who are more junior or code less, daily. Finally, we found a large but not significant interaction effect between use of AI and average hours spent coding per day, suggesting that developers who code more are faster when using AI than those who code less.  

These findings support continued use of and investment in AI-enhanced feature, and invite further research into the mechanisms that lead to differentiated speed gains for developers who code less daily. They also encourage further product development work to explore how we enable all developers to benefit optimally from our AI-enhanced tools.

%% file: paper_sections/02_related_work.tex
\section{Related Work} \label{sec:relatedw}

In recent years, driven by significant improvements in large language models (LLMs), many developer tools have been built upon or incorporated LLMs. GitHub Copilot~\cite{copilot} is one of the earliest LLM-powered developer tools, suggesting code in real time, based on context. Other examples include AlphaCode~\cite{alphacode} from DeepMind, CodeWhisperer~\cite{codewhisperer} from Amazon, Tabnine~\cite{tabnine}, and Cursor~\cite{cursor}. These tools primarily offer code completion, editing, and generation, often with additional chat functionality or context integration to improve code quality and the user experience.  Significant effort has also been invested in developing AI-enhanced software development tools for internal use, such as at Meta~\cite{bader2021} and Google~\cite{garg2022, froemmgen2024, AIsoftEng}, tailored to their developers' workflows and proprietary codebases.

The increasing prevalence of AI-enhanced developer tools has spurred significant research into their benefits and drawbacks across various contexts, including computer science education, open source projects, and industry settings. 
Recent work has explored the potential of these tools to assist students and novice programmers~\cite{prather2024, nguyen2024, kazemitabaar2023}, as well as programmers' perceptions of~\cite{dangelo2024developers, liang2024, chatterjee2024} and trust in~\cite{amoozadeh2023, wang2024trust, brown2024} AI tools. 

A key question surrounding these tools in industry is their potential to enhance developer productivity, leading to improved software quality and reduced development effort and cost. 
While some studies have investigated the \textit{perceived} usefulness of AI-enhanced developer tools (e.g.,~\cite{ziegler2022, bird2023, murillo2023}), quantifying \textit{actual} productivity or speed gains has proven more challenging, due to difficulties in accessing and analyzing unbiased, real-world usage data. 
Moreover, understanding developer productivity itself requires considering multiple factors beyond simple metrics such as lines of code, and existing research attempting to quantify developer productivity---even outside the context of AI tools---highlights the complexities involved.

Despite these challenges, some studies have attempted to measure the impact of AI tools on actual developer productivity. 
For instance, Ziegler et al.~\cite{ziegler2022} analyzed telemetry data to investigate the productivity benefit of GitHub Copilot, and reported over one-fifth of suggestions were accepted by developers, which correlated with their perceived productivity. Similarly, two studies have documented increased throughput for AI in a controlled environment on information-gathering tasks~\cite{nam2024} and in a large, pooled analysis of multiple field experiments, and therefore in enterprise contexts~\cite{cui2024}.
The most relevant study here, however, is \cite{peng2023}, which reports on a randomized controlled experiment comparing developers with and without access to GitHub Copilot, and finds a statistically significant reduction in task completion time.  
However, the latter study's reliance on GitHub Classroom and on participants recruited from Upwork limits its generalizability to real-world, enterprise software development workflows. For instance, about 45\% of their sample participants were self-employed, and about 5\% unemployed; and about 50\% earned less than \$10,000 a year. Our work complements their estimate by studying Google software engineers as they complete a task in a development environment that they know well.

The ``SPACE" framework~\cite{forsgren2021}--encompassing Satisfaction, Performance, Activity, Communication, and Efficiency—--provides a valuable lens for understanding developer productivity holistically, beyond merely lines of code~\cite{weber2024}.  
It emphasizes that productivity extends beyond individual metrics, encompassing multiple dimensions.
Similarly, Murphy-Hill et al.~\cite{murphy2019} highlights the diverse factors influencing developers' self-perceived productivity.
Acknowledging the multifaceted nature of developer productivity, and the potential influence of AI tools on different aspects of productivity, our study focuses on time on task, as encouraged by Hernández-López and colleagues, who found it to be one of the dominant measures of productivity in the software engineering literature~\cite{hernandez2013}. We also investigate the moderating effects of both developer-level and task-level characteristics on the relationship between use of AI and time on task, to minimize the likelihood that what we would attribute to AI would actually be the effect of unobserved characteristics. Our full theoretical framework can be found in \Cref{sec:tcalfhyp}.

%% file: paper_sections/03_dev_at_google.tex
\section{Software Development at Google and Features Included in the Study}\label{sec:atGoogle}

Software development at Google happens in a monolithic source-code repository (or ``monorepo'') called Piper~\cite{froemmgen2024}. Google developers have access to multiple integrated development environments (IDEs), including Cider V, a variant of the VS Code IDE by Microsoft~\cite{vscode}. To commit new code, developers patch changelists (or ``CLs'' for short), the equivalent of ``pull requests'' in the Git ecosystem, and of ``diffs'' at Meta. CLs can be ``patched'' to edit the code.
Once the CL is patched, developers can read and edit code, and changes are tracked and documented as the CL is submitted for review (see~\cite{froemmgen2024} for more details).

\begin{figure}[h!]
\centering
\includegraphics[width=0.9\linewidth]{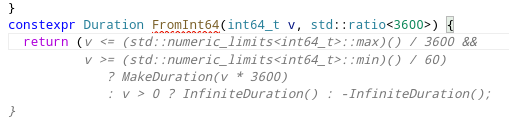}
\caption{AI Code Completion in Cider V. When a user starts typing code, the feature auto-completes the code block in light-grey font based on the context provided. After typing the first line of a new function to be evaluated at compile time, the user starts to type the return logic and the AI Code Completion makes a suggestion based on the entered parameters in light-grey text. Pressing TAB accepts the suggestion.}
\label{fig:codecomplete}
\end{figure}

\begin{figure}[h!]
\centering
\includegraphics[width=0.9\linewidth]{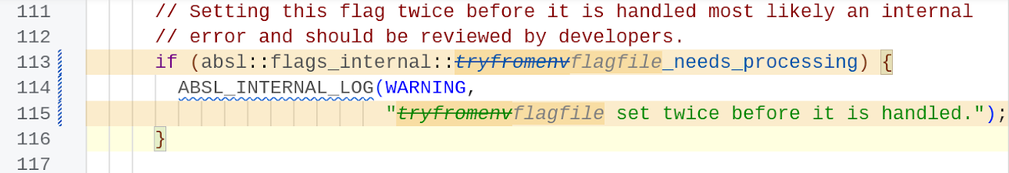}
\caption{Smart Paste feature in Cider V. When a user pastes code, Smart Paste provides an automatic adjustment to the code, then shows the inline diff highlights the removal of tryfromenv (strikethrough) and the insertion of flagfile (italic and lower opacity). The user can accept the adjustment using the established TAB shortcut.}
\label{fig:smartpaste}
\end{figure}

\begin{figure}[h!]
\centering
\includegraphics[width=\linewidth]{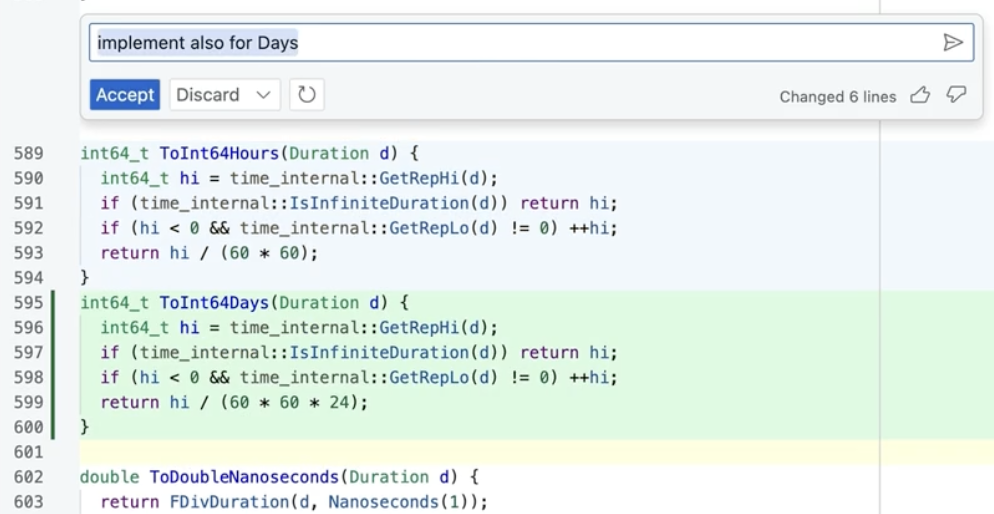}
\caption{Natural language to code feature in Cider V. When users need help to write code, they can move their cursor to where the code can be inserted, and trigger the feature so that a pop-up window hovers above the code, and they can enter their query. The prompt here is ``implement also for days'', and the feature suggests the code that would match the prompt. The user can then review the suggestion and click a button to add the code to their file.}
\label{fig:NL2Code}
\end{figure}

The features we included in this study were all in production and thus available to all Googlers in Cider V at study time. These features were as follows:

\begin{itemize}
    \item \textbf{AI Code Completion.} This feature is a novel, transformer-based hybrid semantic AI code completion that enables single- and multi-line code suggestions as developers type, highlighting the recommendations in light-grey text. Our version~\cite{codecompletion} resembles that built elsewhere in the industry~\cite{dunay2024}. See Figure~\ref{fig:codecomplete} for a visual representation of the feature.
    \item \textbf{Smart Paste.} This feature uses AI to enable context-aware adjustment to code that is pasted from one area to another in the IDE~\cite{smartpaste}. It works in ways that are similar to the well-known copy/paste shortcuts, and displays only suggestions that are high-confidence using light-grey font for the recommendation, and crossing out the text that will be removed. The suggestion is accepted using the ``tab'' key, and ignored otherwise. See Figure~\ref{fig:smartpaste} for a visual overview of the feature.
    \item \textbf{Natural Language to Code.} This feature leverages an AI-assistant trained in Python, Java, Go, C++, TypeScript and JavaScript. To activate, developers move their cursor to (or select) the code area that they want to change. A natural-language to code prompting window enables connection to the model, which makes a suggestion for code fixes (see Figure~\ref{fig:NL2Code} and~\cite{AIsoftEng} for more detail). They can then review the suggestion and either choose to apply or reject it.
\end{itemize}

%% file: paper_sections/04_research_approach.tex
\section{Research Approach} \label{sec:researchA}

To answer our research questions, we designed a randomized controlled trial (RCT). RCTs are a type of scientific experiment designed to assess the effectiveness of a treatment condition by exposing some participants to the treatment (the ``experimental condition'') and some not (the ``control group''). RCTs are often considered an empirical standard for establishing causal links between an intervention and its associated effect, observed empirically, and provide unbiased estimates~\cite{Deaton2018}.

\begin{figure*}[t]
\centering
\includegraphics[width=0.75\linewidth]{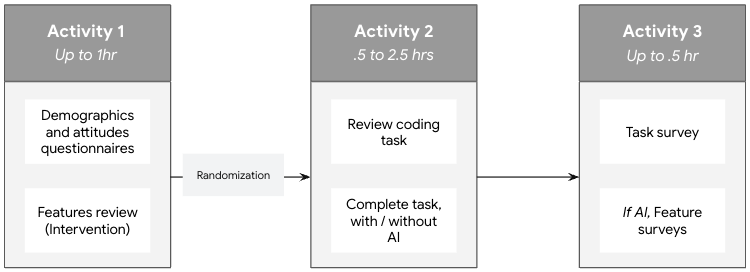}
\caption{Study design: Activities and randomization}
\label{fig:studydes}
\end{figure*}

Our randomized controlled trial was executed as follows (see~\Cref{fig:studydes}). In June and July 2024, full-time software engineers from across Google were recruited via email by a team that was independent from the research team, and were accepted into the study if they met the following basic criteria: they had been working at Google for at least one year, they were proficient in C++, submitted code to Piper, used Cider V as their main IDE, and had some experience with the task domain (see \Cref{sec:atGoogle} above for details on the software development workflow at Google). Once they were recruited, holds were placed on their calendars to book time for the study tasks: a pre-task questionnaire (up to 30 min), an intervention where all participants were trained on the tools included in the study (up to 30 min), completion of a standardized enterprise-grade task (more below in \Cref{sec:task}), and a post-task questionnaire (up to 30 min). Participants were randomly assigned to either the experimental or control group after they had completed both the pre-task questionnaire and the tool training. We elected to train all developers to use our AI-enhanced coding tools to allow us to provide an additional incentive for participation in this extensive trial.

The questionnaires were created after a review of previously-published questions about experience and attitudes towards AI and were tested cognitively following the ``think aloud'' paradigm identified by Beatty and Willis~\cite{beatty2007}: we tested participants’ understanding of the questions, ability to retrieve the expected information and map their answers to the answer scale, as well as their comfort reporting an answer to the question. Participants were randomly assigned to either the AI or no-AI condition after they had completed the pre-task questionnaire and reviewed the AI features. The coding task was identical for the AI and no-AI condition, except for the setup steps, which instructed participants to either enable or disable AI features in their IDE, respectively. As noted above, these features were all in production and commonly available to all developers across the company before the task, but had varied usage rates.

\begin{figure*}
\centering
\includegraphics[width=0.75\linewidth]{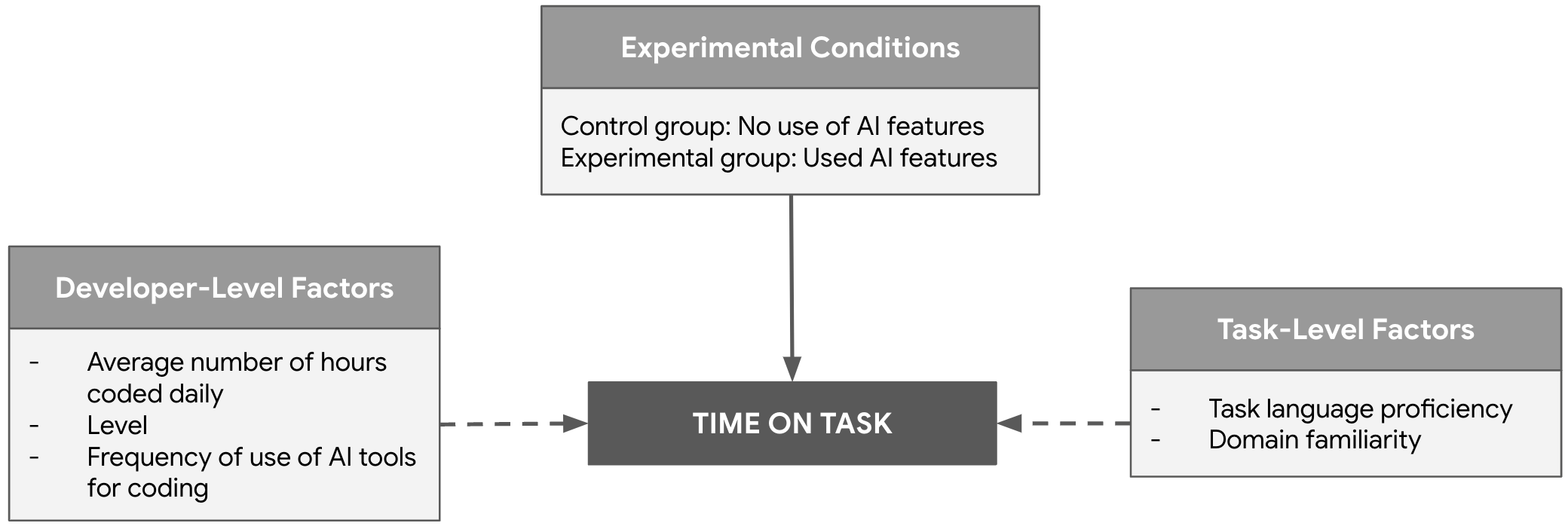}
\caption{Theoretical framework: Factors predicting time on task}
\label{fig:tcalfr}
\end{figure*}

\subsection{Coding task}\label{sec:task}

The coding task was designed by a team of engineers in collaboration with the research team. We started with the task administered by Peng et al.~\cite{peng2023}, and brainstormed similar tasks that would be truly an ``enterprise-grade'' task. To meet this bar, we needed a task that (1) would accurately reflect the type of task that our internal Google developers do on a regular basis \textit{and} (2) would leverage the full range of our developer tools. These two criteria were met with a third, more pragmatic: the task also had to be feasible within the context of this lab study.

The coding task required participants to start from a pre-existing change list (CL) that contained ten files and 474 lines of code, which participants had to edit based on instructions.\footnote{Readers interested in obtaining a copy of the task instructions can reach out to the authors.} Instructions first described how to enable or disable AI features in the IDE, then shared the task goal: to first patch and then edit the CL to implement a new service to log messages from a fake product onto our proprietary internal file storage infrastructure. Participants were given specific instructions for both the server side and log format for messages. To complete the task, participants had to update their build, data structure, and test files to align with the rest of their code and meet task specifications, then build and test their code. The task was considered ``completed'' when all tests passed.

The task was complex enough that it required a fair amount of infrastructure knowledge, code search, code editing, code writing from scratch, and refactoring of test plans, which parallels the typical workflow of developers who need to add a data logging component to a new product feature in our enterprise context. Moreover, since developers were not familiar with the code that they had to patch and edit, and since they had varied levels of familiarity with the task domain (\textit{i.e.} the creation of product-specific data logging infrastructure at Google), they needed to familiarize themselves with the code and with Google standards for such infrastructure, using our internal tools.

To ensure feasibility, this task was pre-tested with 8 software engineers, half with and half without AI tools, and expected to take between 30 minutes and 2.5 hours of development time.

\subsection{Theoretical framework, variables and hypotheses}\label{sec:tcalfhyp}

Since RCTs do not eliminate the need to include control variables in analyses~\cite{Deaton2018}, we created a theoretical framework~\cite{varpio2020} that connects our dependent variable (time spent on task) with the intervention as well as developer- and task-level factors. We created a theoretical framework describing the predicted relationships among multiple components of the developer experience and their impact on time on task (see Figure~\ref{fig:tcalfr}). Theoretical frameworks are a ``logically developed and connected set of concepts and premises … that a researcher creates to scaffold a study''~\cite{varpio2020}. Our framework was influenced by a recent systematic review of the different factors that influence AI adoption~\cite{kelly2023} and the literature on developer productivity~\cite{murphy2019}. 
Using these papers, our internal metrics framework, and a review of the literature on attitudes towards AI (which included~\cite{BERGDAHL2023, coderpad, jetbrains, neudert2020, SCHEPMAN2020}), we generated a list of concepts that we then either instrumentalized through survey questions or telemetry measures. 
As noted above, survey questions were then tested through two rounds of ``think aloud'' cognitive testing~\cite{beatty2007}.

\subsubsection{Dependent variable: Time spent on task}

Our dependent variable was defined as time spent on task. We measured this as the time research participants spent in Cider V working in the study-specific workspace they created at the beginning of the coding task, and on adjoining development surfaces related to coding tasks such as code search and use of debugging tools. The total time spent on task started when they created the study-specific workspace, and ended when their last interaction with the IDE was observed.

\subsubsection{Experimental condition: Use of AI}

Participants were randomly assigned to either the control or experimental group. The control group was asked to complete the task without the help of AI features (identified by ExpCon = 0), while those in the experimental group were asked to enable and then use the three AI coding features available in the IDE (identified by ExpCon = 1). Given the estimates of 56\% speed gain attributable to AI published by Peng et al.~\cite{peng2023} and the 26\% increased throughput measured by Cui et al.~\cite{cui2024}, we generated the following hypothesis:

\textbf{H1: Participants randomly assigned to the AI condition will spend less time on task (main effect)} than those assigned to the control group. We expect our AI tools to decrease time on task, in line with published estimates~\cite{peng2023, cui2024}.

Given that the aforementioned estimates came from a randomized controlled trial and a field experiment, respectively, we expected that the effect attributable to AI to be robust when controlling for other factors, and therefore for the following hypothesis to hold true:

\textbf{H2: Controlling for developer-level and task-level factors, participants randomly assigned to the AI condition will still spend less time on task} than those assigned to the control group. Covariates should not eliminate the main effect of AI on time spent on task in the IDE, but could be independent predictors.

\subsubsection{Moderators at the developer level}

At the developer level, we expected three factors to influence time on task. First, we expected that the average number of hours a developer spends on coding activities, daily (var: AvgProgHrsDay), would influence how quickly they complete the task, and we therefore used this variable as a control variable. We implemented this self-reported variable as an ordinal variable dichotomized into: 0 = zero to four hours on coding activities or 1 = five or more hours spent coding daily. Importantly, current AI features rely on coding expertise to be maximally helpful, since developers using them must still engage in coding, code editing, and code review~\cite{mozannar2024}. Based on this, we generated the following hypothesis:

\textbf{H3: There will be a negative and significant interaction effect between the experimental condition and average daily hours spent coding,} such that those who spend more time coding will be faster with AI tools than those who code less. 

Second, we expected that seniority at the company should influence developer speed on task. In the context of software development, coding expertise accrues with level, but likely asymptotically, since with every level change there is an increase in meeting and supervision loads. We instrumentalized this variable using our internal data table with level information. This variable (var: Level) ranges from 3 to 7, where 3 is early career and 7 is a level just below the first executive level at Google. Seniority is a reflection of not only years of experience, but also of deliberate practice~\cite{ericsson1993}, as it is only with deliberate practice that one reaches expert levels. We therefore used this variable as a control.

Previous research about the interaction between the impact of AI tools and seniority is mixed: some evidence suggests that beginners might benefit more than more experienced developers~\cite{edelman2023}, but other research suggests that professionals might be faster than students when using an AI-enhanced tool for code understanding~\cite{nam2024}. We therefore drafted the following hypothesis:

\textbf{H4: There will be a negative and significant interaction effect between the experimental condition and seniority,} such that more senior people will be faster with AI than more junior people. We make this prediction based on the fact that in pure software engineering terms, developers who work at Google---a top-five tech company---can hardly be claimed to be novices at coding.

Thirdly, we expected that participants’ previous experience with AI tools might enable them to work more quickly with our AI tools. While experience with AI tools in itself might not directly influence someone’s speed on task overall, for both the control and experimental group, we predicted that a developer’s previous experience with AI tools might give them a later advantage when using those tools, compared to less experienced developers. 

We therefore proposed the following hypothesis, positing an interaction effect between experience with AI and speed on task when using AI:

\textbf{H5: There will be a negative and significant interaction effect between the experimental condition and the frequency with which developers use AI coding tools.} Developers who use AI tools more often will be faster with the AI coding tools in the study than those who use them less often, as they have learned to use them.

We instrumentalized this construct as self-reported frequency of AI coding tool use across 13 areas of the software development lifecycle (var: NbrHighFreqAreas). The 13 areas were: writing code, commenting on code, debugging, code explanation, learning a new code base, testing code, preparing change lists, reviewing code, deploying changes, monitoring changes, planning, writing documents, and collaborating with others. We recoded each of the 13 areas into a binary variable, where each area was coded as 0 if AI tools were reported as used once a week or less in that area, or as 1 if they were reportedly used more than once a week. The NbrHighFreqAreas variable is a linear sum of these indicator variables, and ranges from 0 to 13.

\subsubsection{Moderators at the task level}
Among task-level factors, we expected self-reported task language proficiency and self-reported familiarity with the task domain to be predictors of speed on task. We instrumentalized task language proficiency as self-reported use of C++ (the task language) as one of developers’ top-three most-used languages (variable: TopLangCpp). We used an indicator variable where 0 = not in top 3 languages and 1 = in top 3 languages. We instrumentalized domain expertise as self-reported expertise with the task domain (\textit{i.e.} data logging infrastructure at Google) on a 4-point expertise scale where 0 = ‘Learner’ and 3 = ‘Expert’ in the task domain (variable: DataLogExp). We used both as control variables, but posited no interaction effects between task-level factors and use of AI.

\subsection{Analytic approach}

We ran a two-tailed Student t-test ($\alpha = 0.05$) on logged time on task, comparing mean time for participants in the control versus experimental groups.

To explore the robustness of our estimate and test our theoretical model, we ran three follow up linear regressions: one with the complete framework, \textit{i.e.} with the experimental condition, developer-level factors, and task-level factors on logged time on task; one with only the experimental condition and developer-level factors; and one with only the experimental condition and task-level factors.

The full framework can be represented by the following equation:
\vspace{-3pt}
\begin{align*}
LogToT & = \beta_1*ExpCon + \beta_2*AvgProgHrsDay \\
    &    +  \beta_3*Level + \beta_4*NbrHighFreqAreas \\
    &    +  \beta_5*TopLangCpp + \beta_6*DataLogExp +  \epsilon
\end{align*}

Given the logged dependent variable, the effect of a variable $n$ on time on task in percentages is represented by $(1 - exp(\beta_n)) * 100$. The impact of AI on time on task (our ExpCon variable), expressed in percentages, is thus represented by $(1 - exp(\beta_1)) * 100$. We did not use hierarchical linear modeling given the absence of nesting between constructs in our theoretical model.

Finally, to explore interaction effects between experimental condition and developer-level factors, we ran three separate linear regressions with interaction effects, one for each covariate: average hours coded daily, level (seniority), and frequency of use of AI tooling.

To identify the best model, we used adjusted $R^2$ values and model-level $p$ values.

%% file: paper_sections/05_limitations.tex
\section{Limitations}\label{sec:limitations}

We are still very early in the evolution of genAI-based tools for code, and therefore early in the evaluation of their impact. Consequently, we might expect greater impact from these tools over time as the quality and pervasiveness of LLMs and AI-enhanced developer tools increase, and as developers increase their proficiency with them. Documenting the evolution of their impact will therefore be critical, since a “true estimate” of the contribution of AI tools to developer speed will be a moving target. This study is therefore only the beginning of an answer to the impact of AI on developer workflows. 

While this study was adequately powered to test our main effect of the AI feature impact on developer speed, further analyses are needed with larger sample sizes to ascertain interaction effects between use of AI tools and other factors that are known to be relevant to developer speed. A/B tests or field experiments might be best suited to calculate such effects with a much larger sample size.

There may be external validity constraints on our estimates given how participants were all Google employees, a leading global tech company. Findings may therefore not be directly comparable to those obtained from studies with other developer populations, particularly for studies with developers who do not code for a living.

We have attempted to eliminate most sources of bias with our randomization process. We cannot, however, fully eliminate the risk associated with us evaluating our own team’s developer tools. Given the fact that our tools are not available to external developers, however, it was impossible to run the study with developers outside of Google, and also difficult to delegate this work to an outside vendor because of constraints associated with Google's intellectual property.

We strove to make the study environment approximate developers' everyday work environment, and created an enterprise-grade task that aligns closely with what software developers at Google do on the daily. However, experiments like the one we conducted cannot perfectly match the complexity of developers' real-world work, which often has no obvious nor definitive solution. Similarly, the feasibility constraint on our task---which was designed to require no more than 3 hours of work---also limits the potential generalizability of our findings to the broader company or ecosystem levels.

%% file: paper_sections/06_results.tex
\section{Results}
\label{sec:results}

\begin{figure}
\centering
\includegraphics[width=\linewidth, trim={0.3cm 1.5cm 1.5cm 1.5cm},clip]{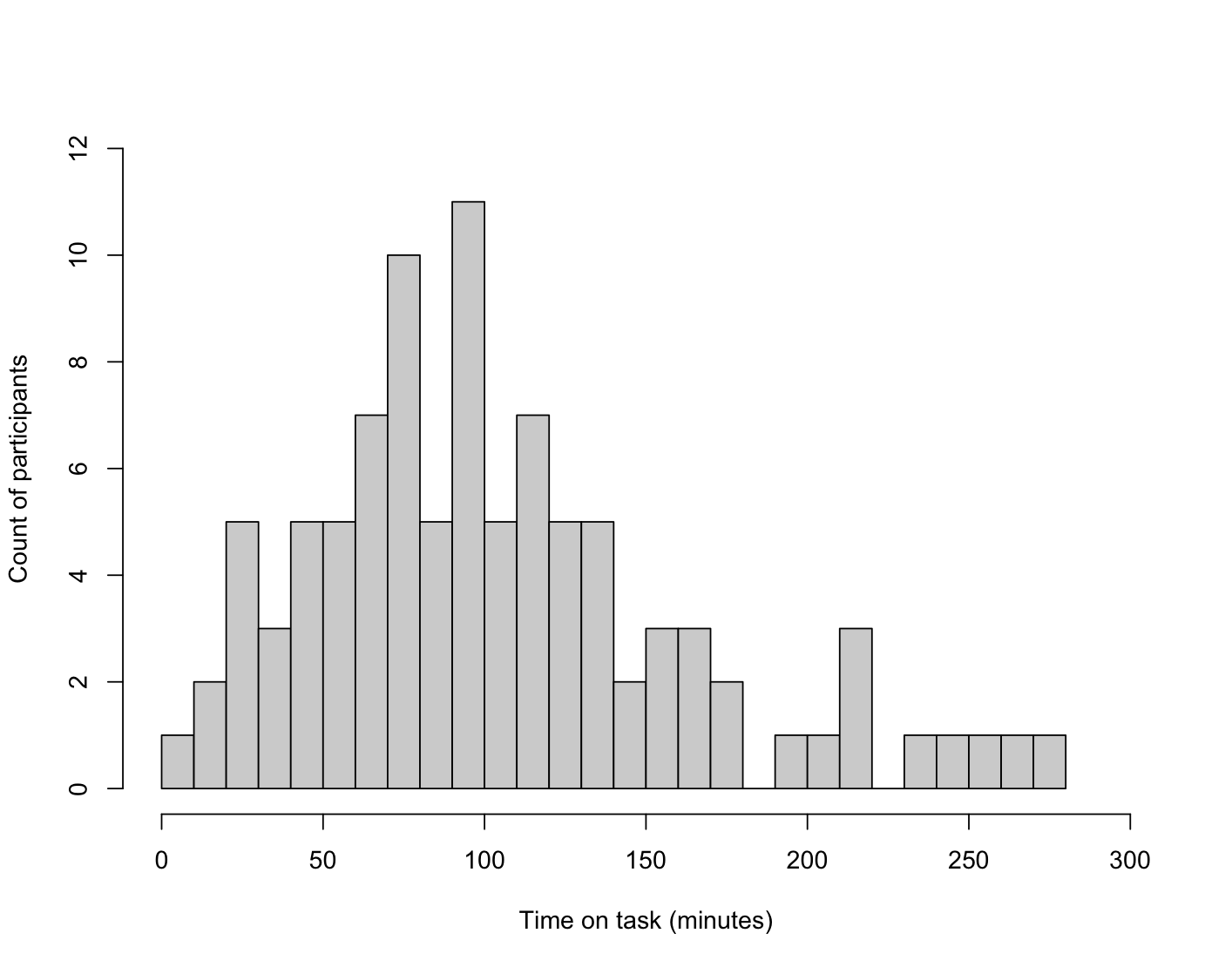}
\caption{Distribution of time spent on study task ($n = 96$}
\label{fig:hist_time}
\end{figure}

\begin{figure}
\centering
\includegraphics[width=\linewidth, trim={0.3cm 1.5cm 1.5cm 1.5cm},clip]{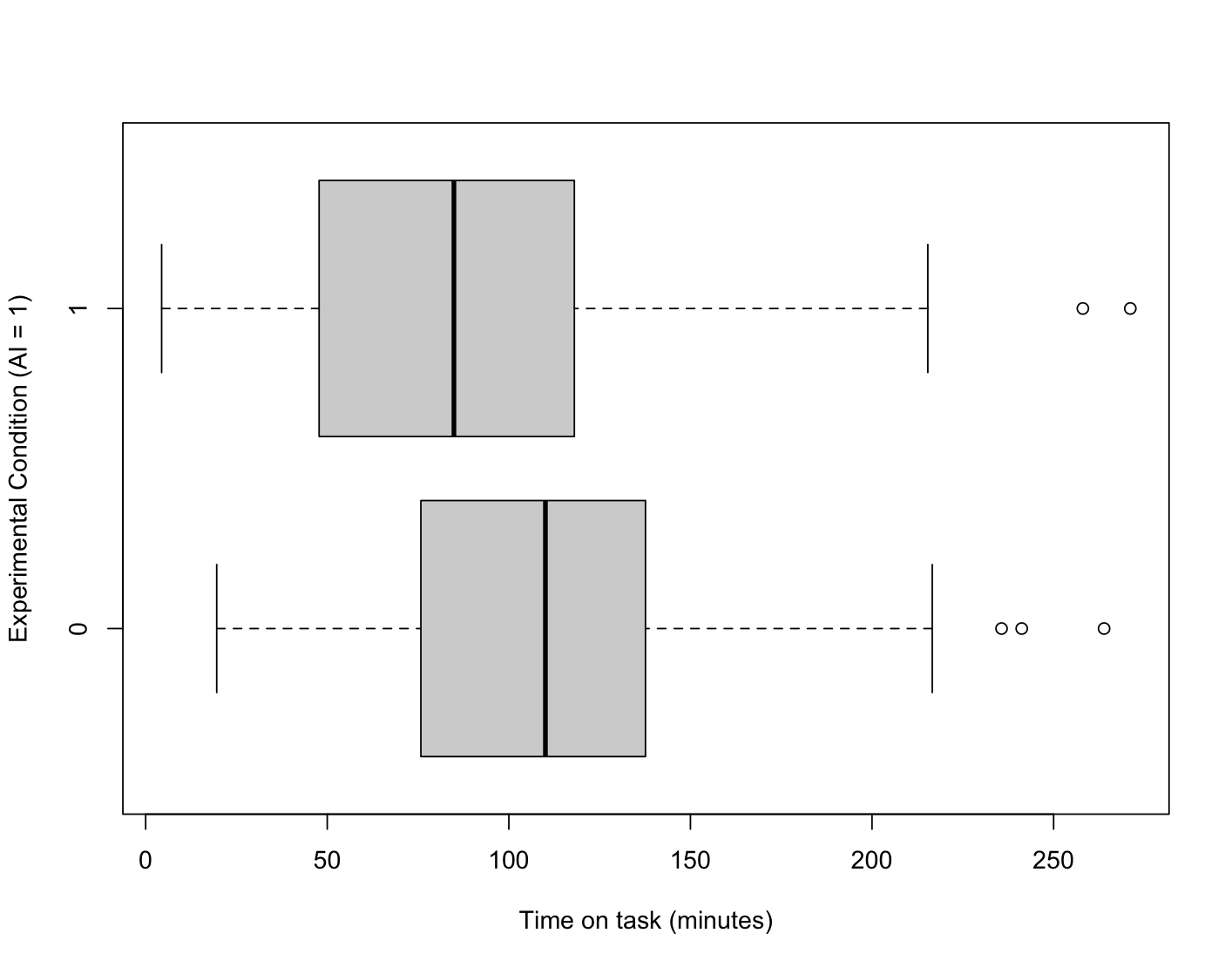}
\caption{Box and whisker plot of time spent on task by experimental condition. The dotted lines represent the first and fourth quartiles, the box represents the interquartile range, and the vertical line in the middle represents the average time on task for each group. Outliers are identified with circles.}
\label{fig:totbyco}
\end{figure}

\begin{table}[t]
\setlength{\tabcolsep}{2pt}
\centering
\caption{Means, standard deviations, and correlations among study variables}
\label{tab:meancorr}
\begin{tabular}{lccccccccc}
\toprule
Variable & Count & M & SD & 1 & 2 & 3 & 4 & 5 & 6 \\
\midrule
1. LogToT           & 96    & 4.46  & 0.69  &       &       &       &       &       &       \\ [1pt]
2. ExpCon           & 96    & 0.50  & 0.50  & -.21* &       &       &       &       &       \\ [1pt]
3. AvgProgHrsDay    & 96    & 0.36  & 0.48  & -.25* & .17   &       &       &       &       \\ [1pt]
4. Level            & 96    & 4.38  & 0.89  & -.14  & -.02  & -.21* &       &       &       \\ [1pt]
5. NbrHighFreqAreas & 96    & 3.30  & 2.73  & -.06  & .12   & .13   & .01   &       &       \\ [1pt]
6. TopLangCpp       & 95    & 0.62  & 0.49  & .03   & .01   & -.01  & -.17  & -.09  &       \\ [1pt]
7. DataLogExp       & 96    & 1.31  & 0.82  & -.01  & .10   & .08   & .07   & .33** & -.23* \\ [1pt]
\bottomrule
\multicolumn{10}{r}{\footnotesize \textit{M} and \textit{SD} are used to represent mean and standard deviation, respectively.} \\
\multicolumn{10}{r}{\footnotesize * indicates $p <$ .05. ** indicates $p <$ .01.} \\
\end{tabular}
\end{table}

\begin{table}[t]
\setlength{\tabcolsep}{4pt}
\centering
\caption{Means, standard deviations, and t-tests results for study covariates, by experimental condition}
\label{tab:ttest}
\begin{tabular}{lcccccc}
\toprule
\multicolumn{1}{l}{} & \multicolumn{2}{c}{Control} & \multicolumn{2}{c}{Experiment} & \multicolumn{2}{c}{} \\
                    & M     & SD    & M     & SD    & Difference & p-value \\
\midrule
3. AvgProgHrsDay    & 0.28 & 0.45 & 0.44 & 0.50 & -1.64 & 0.104 \\ [1pt]
4. Level            & 4.40 & 0.87 & 4.35 & 0.91 & 0.23  & 0.819 \\ [1pt]
5. NbrHighFreqAreas & 2.98 & 2.60 & 3.62 & 2.86 & -1.16 & 0.249 \\ [1pt]
6. TopLangCpp       & 0.62 & 0.49 & 0.63 & 0.49 & -0.08 & 0.937 \\ [1pt]
7. DataLogExp       & 1.23 & 0.88 & 1.40 & 0.77 & -0.99 & 0.325 \\ [1pt]
\bottomrule
\multicolumn{7}{r}{ \textit{M} and \textit{SD} are used to represent mean and standard deviation, respectively.} \\
\multicolumn{7}{r}{T-tests values (Difference) were not statistically significant (all $p >$ .1)} \\
\end{tabular}
\end{table}

A total of 96 participants worked on the task. Of these, 48 were assigned to each condition. Time spent on coding activities related to the study ranged from 4.4 to 271.1 minutes, with a mean of 104.4 minutes (Standard Error (or SE) of the mean $= 60.3$). 93 participants (96.9\%) who started the task also completed it. 
Figure~\ref{fig:hist_time} shows the overall distribution of time spent on task.

\subsection{Pre-analysis sample examination}
Before running the two-tailed student t-test and the linear regressions to test our hypotheses, we first examined whether there exists substantial biases within the data.

First, we used descriptive statistics to examine the characteristics of the sample and the distribution of key variables, as well as correlations among study variables (see Table~\ref{tab:meancorr}). Our dependent variable was significantly and negatively correlated with our experimental condition, and developers’ average number of hours spent on programming tasks, daily. As expected, a negative and significant correlation existed between a person's level and average hours spent coding per day, as well as between C++ expertise (TopLangCpp) and expertise with data logging infrastructure at Google (DataLogExp). A positive and significant correlation was found between expertise with data logging infrastructure at Google (DataLogExp) and the number of areas where developers use AI in their workflows (NbrHighFreqAreas).

We also conducted Welch two-sample t-tests (see Table~\ref{tab:ttest} for details) for all study covariates by experimental condition. As expected given the random assignment of participants to experimental conditions, we found no statistically significant differences between control and experimental groups on our covariates ($p$ values ranged from .10 to .94). 

\subsection{RQ1: Impact of AI tools on developer task speed}
Figure~\ref{fig:totbyco} shows the time distribution by experimental condition in a box and whisker plot. 
On average, developers who were exposed to AI features completed the task in 96 min ($N = 47, SE = 9.3$), compared to 114 min ($N = 46, SE = 8.1$) for developers in the no-AI condition. We conducted a Student’s t-test on log time on task to assess the main effect impact of the experimental condition assignment (AI vs no-AI exposure). Results of this analysis indicated that developers who were in the AI exposure condition (Mean $LogToT = 4.6$) were significantly faster at completing the task ($t(83.6) = 2.11, p = .038$), compared to those who were in the no-AI exposure condition (Mean $LogToT = 4.3$).

\begin{summary}[1: Supported]
% Developers who were in the AI exposure condition were significantly faster at completing the task compared to those who were in the no-AI exposure condition.
Using AI was associated with a shorter time on task ($p <$ 0.05).
\end{summary}

\subsection{RQ2: Impact of covariates on main effect}

Table~\ref{tab:reg} summarizes the regression findings. When controlling for other factors in our theoretical framework, we find that estimates for the impact of the experimental condition on logged time on task range between -0.30 to -0.24 ($\beta_1$ for ExpCon for Models 1-3 in Table~\ref{tab:reg}), suggesting a relatively consistent and negative downward pressure on time on task. Controlling for the developer- and task-level factors described in our theoretical framework (see \Cref{sec:tcalfhyp}), we found that this main effect was robust to covariates, making developers between 21\% and 26\% faster.\footnote{From Model 2, where $\beta_1 = -0.24$, and therefore the effect of AI~$= (1 - exp(\beta_1))* 100 = 21\%$ (with 95\% CI = [-0.51, 0.03]). Similarly, 26\% from Model 3, where $\beta_1 = -0.30$.}
However, our confidence intervals are large, and as a result the estimate is not statistically significant at the $p < 0.05$ level (estimate of experimental condition on best-fit Model 2,  $\beta_1=-0.24;$ ~$95\% CI$ = $[-0.51,0.03]$, $p = 0.086$, NS). Our second hypothesis is therefore only partially supported.

Controlling for experimental condition and other model covariates, average programming hours per day was significantly associated with time spent on task, with participants reporting five or more hours of development work per day being 32\% (\(\beta = -0.38\)) faster than participants reporting zero to four hours of development work per day. Similarly, higher levels of seniority were also associated with a decrease in time on task, with one increased level (from 4 to 5, or 5 to 6, for example) being associated with a 15\% (\(\beta = -0.16\)) decrease in time spent on task.

Other covariates were not significantly associated with time spent on task. As seen in Table~\ref{tab:reg}, the best fit model was Model 2 with developer-level factors only, as evidenced by higher adjusted $R^{2}$ and lower $p$-value. While the total variance accounted for by Model 2 was low, the model was statistically significant ($p = 0.011$), suggesting an adequate fit to our data.

\begin{summary}[2: Partially Supported]

Controlling for other factors, using AI remained associated with a shorter time on task but lost its significance ($p =$ NS).
\end{summary}

\sisetup{
  input-symbols         = {()},
  group-digits          = false,
  table-space-text-post = ***,
  explicit-sign
}

\begin{table}[t]
\setlength{\tabcolsep}{4pt} 
    \centering
    \caption{Summaries of regressions using LogToT as the dependent variable, n = 96. Unstandardized estimates provided, along with standard error of the mean (in parentheses).}
    \label{tab:reg}
    \sisetup{parse-numbers=false}
    \begin{tabular}{lS[table-format=-3.2]S[table-format=-3.2]S[table-format=-3.2]S[table-format=-3.2]}
    \toprule
                    & Complete       & Dev          & Task      &  Interaction \\ 
                    & (1)            & (2)          & (3)       &  (4)         \\ \midrule
    Constant        & 5.36 ***       & 5.41 ***     & 4.55 ***  & 4.66 ***     \\ 
                    & (0.42)         & (0.37)       & (0.19)    & (0.11)       \\ [4pt]
    ExpCon          & -0.25 .        & -0.24 .      & -0.30 *   & -0.14        \\
                    & (0.14)         & (0.14)       & (0.14)    & (0.17)       \\ [4pt]
    AvgProgHrsDay   & -0.38 *        & -0.38 *      &           & -0.16        \\
                    & (0.15)         & (0.15)       &           & (0.22)       \\ [4pt]
    Level           & -0.16 .        & -0.16 *      &           &              \\
                    & (0.08)         & (0.08)       &           &              \\ [4pt]
    NbrHighFreqAreas& -0.01          & 0.00         &           &              \\ 
                    & (0.03)         & (0.03)       &           &              \\ [4pt]
    TopLangCpp      &  0.02          &              & 0.06      &              \\ 
                    & (0.15)         &              & (0.15)    &              \\ [4pt]
    DataLogExp      &  0.05          &              & 0.02      &              \\ 
                    & (0.09)         &              & (0.09)    &              \\ [4pt]
    EC:APHD         &                &              &           & -0.29        \\ 
                    &                &              &           & (0.29)       \\ [4pt] \midrule
    Adj. $R^{2}$    & 0.078          & 0.095        & 0.017     & 0.075        \\
    p value         & 0.040          & 0.011        & 0.209     & 0.018        \\  \bottomrule
    \multicolumn{5}{r}{\footnotesize .: $p<$.1 *: $p<$.05. **: $p<$.01. ***: $p<$0.001} \\ 
    \multicolumn{5}{r}{\footnotesize Note: Estimates are for our logged dependent variable. The effect of each} \\
    \multicolumn{5}{r}{\footnotesize covariate on time on task can be calculated by exponentiating the estimate.} \\
    % \multicolumn{6}{r}{\footnotesize  }\\
\end{tabular}
\end{table}

\subsection{RQ3: Interaction effects between experimental condition and other factors}

Finally, we tested interaction effects between the experimental condition and developer-level factors to identify meaningful patterns whereby some developers with specific characteristics might be differently impacted by the use of AI features. Hypothesis 3 predicted an interaction effect between experimental condition and average hours of programming per day. While none of the covariates were significant, the model itself was significant (see Table 3, Model 4; $p = 0.018$), and the effect size of the coefficient on the interaction term was large and negative (\(\beta = -0.29\)). These data tentatively suggest that those who spend five or more hours on development tasks per day might benefit more from AI than those who spend less time programming, daily.

\vspace{2pt}
\begin{summary}[3: Partially supported]
% There is a negative and significant interaction effect between the use of AI features and the average daily hours spent coding.
Interaction effect was large and negative but not statistically significant, and model fit for Model 4 was significant (Adjusted $R^2 = 0.075; p = 0.018$).
\end{summary}
\vspace{5pt}

The interaction between experimental condition and seniority was not statistically significant (model not shown; $p = 0.706$), but it was negative, suggesting that participants in our sample with increased levels of seniority were faster with AI than those with lower levels of seniority. Similarly, the interaction effect between experimental condition and higher frequency of usage of AI tools was negative but not statistically significant (model not shown; $p = 0.235$). These results reject both Hypotheses 4 and 5.

\begin{summary}[4 and 5: Rejected]
% There is no interaction effects between the use of AI features and the seniority or the frequency with which developers use AI coding tools.
Interaction effects between the use of AI features and the seniority or the frequency with which developers use AI coding tools were both negative but not statistically significant. 
\end{summary}

%% file: paper_sections/07_discussion_implications.tex
\section{Discussion and implications}\label{sec:discuss}

In this study, we aimed to quantify the impact of using AI coding features on the time developers take to complete an enterprise-grade, standardized task.

Our analyses found that developers who used AI features were statistically significantly faster than those who did not. This suggests that the AI-enhanced features included in the study (see \Cref{sec:atGoogle}) do indeed make developers faster, and thus supports our first hypothesis ($t(83.6) = 2.11, p = .038$). 

When controlling for other factors in our theoretical framework, we obtained a similar effect size for AI, but the effect did not meet the $p < .05$ significance threshold. At this stage, we estimate a roughly 21\% increase in development speed attributable to AI, controlling for other important predictors (\textit{i.e.} the effect attributable to AI in our best-fit model, Model 2).
This estimate is significantly smaller than the 56\% estimate shared by Peng and colleagues about GitHub Copilot~\cite{peng2023}, but aligns with the 26\% productivity increase attributed to Copilot by Cui and colleagues~\cite{cui2024}. 

The difference between our estimate of developer speed gained through AI and that published in \cite{peng2023} is likely attributable to two main factors. There may be important differences between our suite of AI-enhanced tools and those used by GitHub Copilot. More likely, however, the difference is caused by differences in the underlying populations we recruited from. Indeed, Peng et al.~\cite{peng2023} recruited from Upwork (a freelancing platform), while we only sampled full-time Googlers. Their sample is therefore likely much more diverse than ours when it comes to coding experience and expertise. It is no surprise, then, that our estimate is closer to that offered by~\cite{cui2024}, which focused on enterprise users.

When it comes to the impact of the time participants spend on coding tasks daily on their speed on our task, our data suggest that developers who code more hours per day may be faster with AI tools than those who code less. While our interaction effect was not significant ($\beta = -0.29$), the effect size was large, negative, and the model itself was significant ($p = 0.018$). This might be because our current AI tools require that developers spend a lot of time verifying and editing code generated by AI, which may confer an advantage to developers who spend more time working with code.

Importantly for the future of AI-enhanced developer tools as products serving very diverse subsets of developers, our data suggest that more senior developers, as defined by their level, may work \textit{even faster} with our AI tools. This is in contrast with some findings in the literature, which suggest that more junior developers stand to benefit more from AI~\cite{edelman2023}, but aligns with other work by Nam and colleagues, who found that coding experience might amplify the value gained from AI~\cite{nam2024}. We believe that this may be because AI is not yet able to close a skill gap when applied to complex tasks such as the one on which we tested our participants. To bridge skill gaps, more research and development work will be needed---along the lines of personalization, perhaps---to make more junior developers even faster with AI.

While the exact mechanisms connecting hours spent coding and seniority with increased task velocity with AI tools are still unclear, approaches like the one proposed by Mozannar and colleagues~\cite{mozannar2024} are likely to be a viable path forward in identifying what is truly going on at the intersection of expertise and speed with AI. Indeed, the code generated by our AI features in such a context is still complex; users might therefore still need to understand highly-nuanced, large-scale systems to be effective while using these features. More detailed, logs-based analyses might circumvent the issues we faced with statistical significance within our trial data.

We found no statistically significant evidence that more frequent use of AI coding tools in participants’ software development workflow increased their speed when using AI during our task. This might be attributable to one of four reasons, beyond inadequate statistical power to detect such a difference with the variance on our outcome variable. First, our tools might have a ceiling effect and might not enable an ``expert'' mode that would support greater velocity for those who have reached that expert level, compared to newcomers. Second, our tools might have a low usability floor: benefits might accrue very quickly if there is not a high bar to entry. Third, our tools might not be cohesive enough to support knowledge transfer from one set of tools to another across the development workflow. Finally, while some of our features such as Smart Paste and AI Code Completion will be familiar to developers, others require a steeper learning curve: Natural Language to Code in particular requires familiarity with LLMs and prompting, skills which might not be universal at this time, despite the training we offered participants ahead of task completion (see also~\cite{weber2024}). Here again we find product implications: while we should celebrate the low floor to usability of some of our tools, we can likely do better with both giving access to the more complex features for newcomers, and with giving access to more advanced features or ``expert modes’’ to those who want to go even faster.

There may also be profound educational implications to these technologies and our findings, in particular: How might we up-level those who do not yet have the knowledge and skills to leverage AI successfully, especially with our more complex tools? How might we safeguard against deskilling, and ensure that our developers retain the critical skills---critical thinking, systems knowledge, coding knowledge---that they need to be successful as developers over the course of their careers? And how might we convince the most resistant and least trusting~\cite{brown2024} among our workforce to embrace this technology, so that they are not left behind and can also benefit from AI? These concerns over knowledge, equity, and quality might be particularly important at Google scale, in enterprise contexts, and for highly complex tasks, where the full power of AI might be manifested.

Finally, the insignificance of task-level factors as predictors of task speed is worth further exploration. It is unclear why expertise with the task domain and coding language were not significant predictors of time spent on task. Our sample might have been too homogeneous to detect such effects. Further research here might also be needed.

%% file: paper_sections/08_conclusion.tex
\section{Conclusion}\label{sec:concl}
While our team is bullish on the speed gains that will be realized by AI-enhanced coding tools, questions of equity are not settled by our study, and questions about the impact of AI on code quality were not explored.
Much more research and development is required to explore and potentially remedy the differentiated impact that AI-enhanced tools might have on people of diverse seniority levels. While our study hinted to the positive impact of AI tools on more senior developers, other research has found contradictory effects~\cite{nam2024,edelman2023}. The future might see the targeting of different tools to different types of developers, or learn directly from developers what is optimal for their own specific needs and developmental stages, and provide personalized experiences.

Importantly, there is a growing literature suggesting that while coding assistants might be increasing the total number of code contributions~\cite{cui2024,yeverechyahu2024}, AI might be lowering code quality at the ecosystem level~\cite{gitclear2023}, might increase code churn~\cite{imai2022}, and might not have reached a quality bar---either on the model or UI side---that improves task completion rates~\cite{vaithilingam2022, Zhang2023}. Careful research and development work that balances the sometimes-contradictory incentives to move fast and guarantee high-quality code will ensure the long-term success of AI in the developer space, especially if such work focuses on where developers might optimally trust AI and therefore want to delegate more work~\cite{dangelo2024developers,sergeyuk2024using}.

Finally: since this research provides only one data point, gleaned from one lab study and at a very specific point in the history of AI-enhanced developer tools and since questions about return on investment in AI and comparison across developer suites will continue to arise, it is critical for the field to keep researching and publishing these estimates and pull the field forward.

%% file: paper_sections/09_acknowledgements.tex
\section{Acknowledgements}\label{sec:ack}
The authors wish to acknowledge the contributions of the following people: Don Eriko Anselmo, Paige Bailey, Rico Cruz, Sarah D'Angelo, Tao Dong, Madhura Dudhgaonkar, Brett Durrett, Mona El Mahdy, Mike Giardina, Shivani Govil, Matthew Hughes, Joshua Katz, Min Kim, Angelo Luo, Dan Mcclary, Ryan McGarry, Cody Miller, Kristof Molnar, Ambar Murilo, Nicole Ortiz, Sara Ortloff, Mauli Pandey, Robin Savinar, Johann Scheidt, Niranjan Tulpule, and all study participants.

%% file: main.bbl
\begin{thebibliography}{10}
\providecommand{\url}[1]{#1}
\csname url@samestyle\endcsname
\providecommand{\newblock}{\relax}
\providecommand{\bibinfo}[2]{#2}
\providecommand{\BIBentrySTDinterwordspacing}{\spaceskip=0pt\relax}
\providecommand{\BIBentryALTinterwordstretchfactor}{4}
\providecommand{\BIBentryALTinterwordspacing}{\spaceskip=\fontdimen2\font plus
\BIBentryALTinterwordstretchfactor\fontdimen3\font minus
  \fontdimen4\font\relax}
\providecommand{\BIBforeignlanguage}[2]{{%
\expandafter\ifx\csname l@#1\endcsname\relax
\typeout{** WARNING: IEEEtran.bst: No hyphenation pattern has been}%
\typeout{** loaded for the language `#1'. Using the pattern for}%
\typeout{** the default language instead.}%
\else
\language=\csname l@#1\endcsname
\fi
#2}}
\providecommand{\BIBdecl}{\relax}
\BIBdecl

\bibitem{vaswani2017}
\BIBentryALTinterwordspacing
A.~Vaswani, N.~Shazeer, N.~Parmar, J.~Uszkoreit, L.~Jones, A.~N. Gomez,
  L.~Kaiser, and I.~Polosukhin, ``Attention is all you need,'' 2023. [Online].
  Available: \url{https://arxiv.org/abs/1706.03762}
\BIBentrySTDinterwordspacing

\bibitem{cnn-2023-chatRevolution}
C.~Thorbecke, ``A year after {ChatGPT}'s release, the {AI} revolution is just
  beginning,''
  \url{https://www.cnn.com/2023/11/30/tech/chatgpt-openai-revolution-one-year/index.html},
  2023, accessed: 2024-06-24.

\bibitem{copilot}
``Copilot,'' \url{https://github.com/features/copilot}, accessed: 2024-10-07.

\bibitem{alphacode}
``Alphacode,'' \url{https://alphacode.deepmind.com/}, accessed: 2024-10-07.

\bibitem{codewhisperer}
``Codewhisperer,''
  \url{https://docs.aws.amazon.com/codewhisperer/latest/userguide/what-is-cwspr.html},
  accessed: 2024-10-07.

\bibitem{tabnine}
``Tabnine,'' \url{https://www.tabnine.com/}, accessed: 2024-10-07.

\bibitem{cursor}
``Cursor,'' \url{https://www.cursor.com/}, accessed: 2024-10-07.

\bibitem{bader2021}
J.~Bader, S.~S. Kim, F.~S. Luan, S.~Chandra, and E.~Meijer, ``Ai in software
  engineering at facebook,'' \emph{IEEE Software}, vol.~38, no.~4, pp. 52--61,
  2021.

\bibitem{garg2022}
S.~Garg, R.~Z. Moghaddam, C.~B. Clement, N.~Sundaresan, and C.~Wu,
  ``Deepdev-perf: a deep learning-based approach for improving software
  performance,'' in \emph{Proceedings of the 30th ACM Joint European Software
  Engineering Conference and Symposium on the Foundations of Software
  Engineering}, 2022, p. 948–958.

\bibitem{froemmgen2024}
A.~Froemmgen, J.~Austin, P.~Choy, N.~Ghelani, L.~Kharatyan, G.~Surita,
  E.~Khrapko, P.~Lamblin, P.-A. Manzagol, M.~Revaj, M.~Tabachnyk, D.~Tarlow,
  K.~Villela, D.~Zheng, S.~Chandra, and P.~Maniatis, ``Resolving code review
  comments with machine learning,'' in \emph{Proceedings of the 46th
  International Conference on Software Engineering: Software Engineering in
  Practice}, 2024, p. 204–215.

\bibitem{AIsoftEng}
S.~Chandra and M.~Tabachnyk, ``{AI} in software engineering at google: Progress
  and the path ahead,''
  \url{https://research.google/blog/ai-in-software-engineering-at-google-progress-and-the-path-ahead/},
  Google, 2024, accessed: 2024-07-10.

\bibitem{peng2023}
S.~Peng, E.~Kalliamvakou, P.~Cihon, and M.~Demirer, ``The impact of ai on
  developer productivity: Evidence from github copilot,'' \emph{arXiv preprint
  arXiv:2302.06590}, 2023.

\bibitem{cui2024}
Z.~K. Cui, M.~Demirer, S.~Jaffe, L.~Musolff, S.~Peng, and T.~Salz, ``The
  effects of generative ai on high skilled work: Evidence from three field
  experiments with software developers,'' \emph{Available at SSRN}, 2024.

\bibitem{ziegler2022}
A.~Ziegler, E.~Kalliamvakou, X.~A. Li, A.~Rice, D.~Rifkin, S.~Simister,
  G.~Sittampalam, and E.~Aftandilian, ``Productivity assessment of neural code
  completion,'' in \emph{Proceedings of the 6th ACM SIGPLAN International
  Symposium on Machine Programming}, 2022, pp. 21--29.

\bibitem{nam2024}
D.~Nam, A.~Macvean, V.~Hellendoorn, B.~Vasilescu, and B.~Myers, ``Using an llm
  to help with code understanding,'' in \emph{Proceedings of the IEEE/ACM 46th
  International Conference on Software Engineering}, 2024.

\bibitem{vaithilingam2022}
P.~Vaithilingam, T.~Zhang, and E.~L. Glassman, ``Expectation vs. experience:
  Evaluating the usability of code generation tools powered by large language
  models,'' in \emph{Extended Abstracts of the 2022 CHI Conference on Human
  Factors in Computing Systems}, 2022.

\bibitem{imai2022}
S.~Imai, ``Is github copilot a substitute for human pair-programming? an
  empirical study,'' in \emph{2022 IEEE/ACM 44th International Conference on
  Software Engineering: Companion Proceedings (ICSE-Companion)}, 2022, pp.
  319--321.

\bibitem{varpio2020}
L.~Varpio, E.~Paradis, S.~Uijtdehaage, and M.~Young, ``The distinctions between
  theory, theoretical framework, and conceptual framework,'' \emph{Academic
  medicine}, vol.~95, no.~7, pp. 989--994, 2020.

\bibitem{prather2024}
J.~Prather, B.~N. Reeves, P.~Denny, B.~A. Becker, J.~Leinonen,
  A.~Luxton-Reilly, G.~Powell, J.~Finnie-Ansley, and E.~A. Santos, ````it’s
  weird that it knows what i want'': Usability and interactions with copilot
  for novice programmers,'' \emph{ACM Trans. Comput.-Hum. Interact.}, vol.~31,
  no.~1, Nov. 2023.

\bibitem{nguyen2024}
S.~Nguyen, H.~M. Babe, Y.~Zi, A.~Guha, C.~J. Anderson, and M.~Q. Feldman, ``How
  beginning programmers and code llms (mis)read each other,'' in
  \emph{Proceedings of the 2024 CHI Conference on Human Factors in Computing
  Systems}, 2024.

\bibitem{kazemitabaar2023}
M.~Kazemitabaar, J.~Chow, C.~K.~T. Ma, B.~J. Ericson, D.~Weintrop, and
  T.~Grossman, ``Studying the effect of ai code generators on supporting novice
  learners in introductory programming,'' in \emph{Proceedings of the 2023 CHI
  Conference on Human Factors in Computing Systems}, 2023, pp. 1--23.

\bibitem{dangelo2024developers}
S.~D’Angelo, A.~Murillo, S.~Chandra, and A.~Macvean, ``What do developers
  want from ai?'' \emph{IEEE Software}, vol.~41, no.~3, pp. 11--15, 2024.

\bibitem{liang2024}
J.~T. Liang, C.~Yang, and B.~A. Myers, ``A large-scale survey on the usability
  of ai programming assistants: Successes and challenges,'' in
  \emph{Proceedings of the IEEE/ACM 46th International Conference on Software
  Engineering}, 2024.

\bibitem{chatterjee2024}
S.~Chatterjee, C.~L. Liu, G.~Rowland, and T.~Hogarth, ``The impact of ai tool
  on engineering at anz bank an emperical study on github copilot within
  coporate environment,'' \emph{arXiv preprint arXiv:2402.05636}, 2024.

\bibitem{amoozadeh2023}
M.~Amoozadeh, D.~Daniels, S.~Chen, D.~Nam, A.~Kumar, M.~Hilton, M.~A. Alipour,
  and S.~S. Ragavan, ``Towards characterizing trust in generative artificial
  intelligence among students,'' in \emph{Proceedings of the 2023 ACM
  Conference on International Computing Education Research - Volume 2}, 2023,
  p. 3–4.

\bibitem{wang2024trust}
R.~Wang, R.~Cheng, D.~Ford, and T.~Zimmermann, ``Investigating and designing
  for trust in ai-powered code generation tools,'' in \emph{Proceedings of the
  2024 ACM Conference on Fairness, Accountability, and Transparency}, 2024, p.
  1475–1493.

\bibitem{brown2024}
A.~Brown, S.~D'Angelo, A.~Murillo, C.~Jaspan, and C.~Green, ``Identifying the
  factors that influence trust in ai code completion,'' in \emph{Proceedings of
  the 1st ACM International Conference on AI-Powered Software}, 2024, pp. 1--9.

\bibitem{bird2023}
C.~Bird, D.~Ford, T.~Zimmermann, N.~Forsgren, E.~Kalliamvakou, T.~Lowdermilk,
  and I.~Gazit, ``Taking flight with copilot,'' \emph{Commun. ACM}, vol.~66,
  no.~6, p. 56–62, May 2023.

\bibitem{murillo2023}
A.~Murillo, A.~Macvean, I.~Chu, Q.~Madison, and S.~D’Angelo, ``“if it’s
  what i wanted that’s great, but if it’s not, i just wasted time”:
  Unpacking the perceived costs/benefits of ml enhanced developer tooling.'' in
  \emph{Proc. Deep Learning 4 Code Workshop}, 2023, pp. 1--9.

\bibitem{forsgren2021}
N.~Forsgren, M.-A. Storey, C.~Maddila, T.~Zimmermann, B.~Houck, and J.~Butler,
  ``The space of developer productivity: There's more to it than you think.''
  \emph{Queue}, vol.~19, no.~1, p. 20–48, mar 2021.

\bibitem{weber2024}
T.~Weber, M.~Brandmaier, A.~Schmidt, and S.~Mayer, ``Significant productivity
  gains through programming with large language models,'' \emph{Proceedings of
  the ACM on Human-Computer Interaction}, vol.~8, no. EICS, pp. 1--29, 2024.

\bibitem{murphy2019}
E.~Murphy-Hill, C.~Jaspan, C.~Sadowski, D.~Shepherd, M.~Phillips, C.~Winter,
  A.~Knight, E.~Smith, and M.~Jorde, ``What predicts software developers’
  productivity?'' \emph{IEEE Transactions on Software Engineering}, vol.~47,
  no.~3, pp. 582--594, 2019.

\bibitem{hernandez2013}
A.~Hern{\'a}ndez-L{\'o}pez, R.~Colomo-Palacios, and {\'A}.~Garc{\'\i}a-Crespo,
  ``Software engineering job productivity—a systematic review,''
  \emph{International Journal of Software Engineering and Knowledge
  Engineering}, vol.~23, no.~03, pp. 387--406, 2013.

\bibitem{vscode}
V.~S. Code, ``Visual studio code,'' \url{https://code.visualstudio.com/},
  accessed: 2024-10-03.

\bibitem{codecompletion}
M.~Tabachnyk, ``Ml-enhanced code completion improves developer productivity,''
  \url{https://research.google/blog/ml-enhanced-code-completion-improves-developer-productivity/},
  2024, accessed: 2024-10-03.

\bibitem{dunay2024}
O.~Dunay, D.~Cheng, A.~Tait, P.~Thakkar, P.~C. Rigby, A.~Chiu, I.~Ahmad,
  A.~Ganesan, C.~Maddila, V.~Murali \emph{et~al.}, ``Multi-line ai-assisted
  code authoring,'' in \emph{Companion Proceedings of the 32nd ACM
  International Conference on the Foundations of Software Engineering}, 2024,
  pp. 150--160.

\bibitem{smartpaste}
S.~Forte, ``Smart paste for context-aware adjustments to pasted code,''
  \url{https://research.google/blog/smart-paste-for-context-aware-adjustments-to-pasted-code/},
  2024, accessed: 2024-10-03.

\bibitem{Deaton2018}
A.~Deaton and N.~Cartwright, ``Understanding and misunderstanding randomized
  controlled trials,'' \emph{Social Science \& Medicine}, vol. 210, pp. 2--21,
  2018, randomized Controlled Trials and Evidence-based Policy: A
  Multidisciplinary Dialogue.

\bibitem{beatty2007}
P.~C. Beatty and G.~B. Willis, ``Research synthesis: The practice of cognitive
  interviewing,'' \emph{Public Opinion Quarterly}, vol.~71, no.~2, pp.
  287--311, 05 2007.

\bibitem{kelly2023}
S.~Kelly, S.-A. Kaye, and O.~Oviedo-Trespalacios, ``What factors contribute to
  the acceptance of artificial intelligence? a systematic review,''
  \emph{Telematics and Informatics}, vol.~77, p. 101925, 2023.

\bibitem{BERGDAHL2023}
J.~Bergdahl, R.~Latikka, M.~Celuch, I.~Savolainen, E.~{Soares Mantere},
  N.~Savela, and A.~Oksanen, ``Self-determination and attitudes toward
  artificial intelligence: Cross-national and longitudinal perspectives,''
  \emph{Telematics and Informatics}, vol.~82, p. 102013, 2023.

\bibitem{coderpad}
CoderPad, ``Coderpad state of tech hiring,''
  \url{https://coderpad.io/survey-reports/coderpad-and-codingame-state-of-tech-hiring-2024/},
  2024, accessed: 2024-07-10.

\bibitem{jetbrains}
J.~T.~I. Lab, ``Jetbrains the state of developer ecosystem 2023,''
  \url{https://www.jetbrains.com/lp/devecosystem-2023/}, JetBrains, 2023,
  accessed: 2024-07-10.

\bibitem{neudert2020}
L.-M. Neudert, A.~Knuutila, and P.~N. Howard, ``Global attitudes towards ai,
  machine learning \& automated decision making–implications for involving
  artificial intelligence in public service and good governance,''
  \url{https://perma.cc/6PB6-X56B}, 2023, accessed: 2024-07-10.

\bibitem{SCHEPMAN2020}
A.~Schepman and P.~Rodway, ``Initial validation of the general attitudes
  towards artificial intelligence scale,'' \emph{Computers in Human Behavior
  Reports}, vol.~1, p. 100014, 2020.

\bibitem{mozannar2024}
H.~Mozannar, G.~Bansal, A.~Fourney, and E.~Horvitz, ``Reading between the
  lines: Modeling user behavior and costs in ai-assisted programming,'' in
  \emph{Proceedings of the 2024 CHI Conference on Human Factors in Computing
  Systems}, New York, NY, USA, 2024.

\bibitem{ericsson1993}
K.~A. Ericsson, R.~T. Krampe, and C.~Tesch-R{\"o}mer, ``The role of deliberate
  practice in the acquisition of expert performance.'' \emph{Psychological
  review}, vol. 100, no.~3, p. 363, 1993.

\bibitem{edelman2023}
B.~G. Edelman, J.~Bono, S.~Peng, R.~Rodriguez, and S.~Ho, ``Randomized
  controlled trial for microsoft security copilot,'' \emph{Available at SSRN
  4648700}, 2023.

\bibitem{yeverechyahu2024}
D.~Yeverechyahu, R.~Mayya, and G.~Oestreicher-Singer, ``The impact of large
  language models on open-source innovation: Evidence from github copilot,''
  \emph{arXiv preprint arXiv:2409.08379}, 2024.

\bibitem{gitclear2023}
W.~Harding and M.~Kloster, ``Coding on copilot: 2023 data suggests downward
  pressure on code quality,''
  \url{https://www.gitclear.com/coding_on_copilot_data_shows_ais_downward_pressure_on_code_quality/},
  2024, accessed: 2024-07-10.

\bibitem{Zhang2023}
B.~Zhang, P.~Liang, X.~Zhou, A.~Ahmad, and M.~Waseem, ``Practices and
  challenges of using github copilot: An empirical study,'' in
  \emph{Proceedings of the 35th International Conference on Software
  Engineering and Knowledge Engineering}, vol. 2023, jul 2023, p. 124–129.

\bibitem{sergeyuk2024using}
A.~Sergeyuk, Y.~Golubev, T.~Bryksin, and I.~Ahmed, ``Using ai-based coding
  assistants in practice: State of affairs, perceptions, and ways forward,''
  \emph{arXiv preprint arXiv:2406.07765}, 2024.

\end{thebibliography}
